# Prediction of Dementia-related Agitation Using Multivariate Ambient Environmental Time-series Data

Nutta Homdee
*Link Lab*
Department of Electrical and Computer Engineering
University of Virginia
Charlottesville, VA
nh4ar@virginia.edu

John Lach
*Link Lab*
Department of Electrical and Computer Engineering
University of Virginia
and
George Washington University
jlach@gwu.edu

*Abstract*—Dementia-related agitation causes high stress for dementia caregivers (CG) and to persons with dementia (PWD). Current clinical research suggests that dementia agitation can be affected or triggered by the ambient environment and other contextual factors. In this study, we evaluate this hypothesis through an analysis of ambient environmental data collected with a remote sensing system deployed in the homes of PWDs and their CGs. Furthermore, we determine if the occurrence of dementia-related agitation can be *predicted* from ambient environmental data, creating the potential for agitation to be prevented via the environmental alteration. These collected data are used to learn the environmental patterns using a predictive model approach. The agitation labels, used in model training, are provided by the CGs living with the PWDs. The results of the agitation prediction model evaluation suggest that ambient environment can be used as predictors for upcoming dementia-related agitation. We also observed that environmental triggers for agitation are PWD-specific. Future opportunities and techniques used to understand triggers for dementia agitation are also discussed.

I. INTRODUCTION

Persons with dementia (PWD) suffer from declined memory or other thinking skills which reduce the PWD's ability to perform everyday activities [1]. The difficult daily lives of the PWD can cause the patient to express agitated behavior such as verbal outbursts, or aggressive motor behaviors [2]. Dementia agitation from the PWD, if not prevented at the early stages, could cause a high burden and stress to the dementia caregiver (CG) [3], who is often the patient's spouse, or other relatives who reside in the same residence. This CG burden associated with dementia agitation has been reported to be one of the principal factors prompting the institutionalization of community-dwelling PWD [3]. Considering the expanding aging population [4], it is important to ensure that PWDs are able to remain in the community as long as possible. Thus, we need effective agitation prevention strategies by understanding the triggers of dementia-related agitation.

The occurrence of agitation can be unpredictable or be affected by the ambient environment and other contexts [2], but such factors can be dynamic and PWD-specific. There are studies which show that dementia agitation can be caused by the ambient environment. A study [5] from the college of nursing, University of Wisconsin-Milwaukee suggests that dementia agitation has a significant correlation with the level of sound in the environment (in this case, nursing homes). Another study [6] observed the frequency of agitation occurrences in different light intensities environment; the study shows that high-intensity bright light can affect restlessness behavior in institutionalized older adults with dementia.

From reports showing the correlation between dementia agitation and the ambient environment of the PWD, this work aims to evaluate the possibility that ambient environmental time-series data can be used to predict a potential upcoming agitation. Using ambient environmental data to predict dementia agitation could have various potential benefits such as: notify CG of upcoming agitations, or study the environmental triggers of agitation and used as an agitation intervention. The main contributions of this work are:

- The evaluation of the ability to predict upcoming dementia agitations using a novel predictor - ambient environmental data.
- Data processing framework and predictive modeling techniques to predict dementia agitation from the ambient environment.

This work is supported by the National Science Foundation under grant IIS-1418622

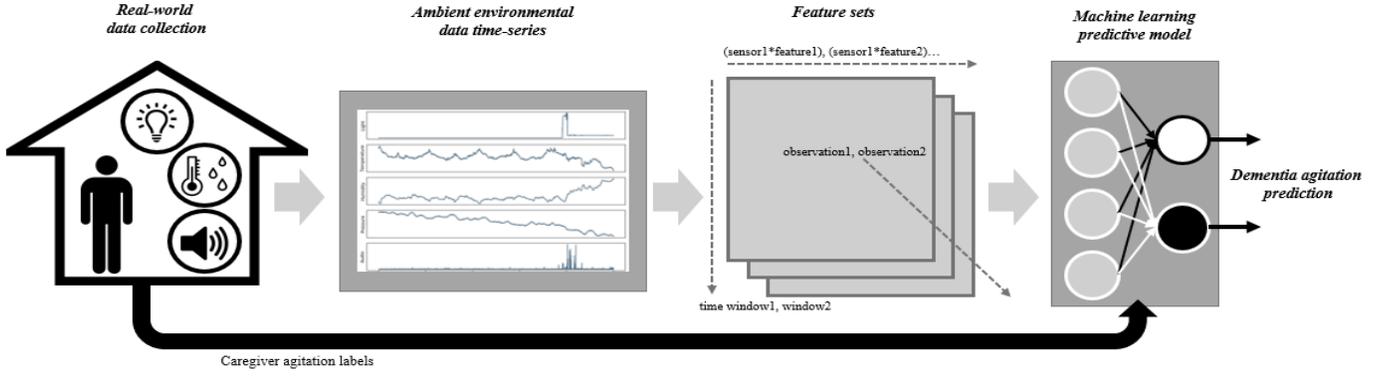

Fig. 1. An illustration of the proposed methodology for predicting upcoming dementia agitation using environmental data time-series. This includes: data collection, data processing framework, and agitation prediction model.

## II. METHODOLOGY

This section explains: (1) the data collection of the ambient environment and agitation labels from CG reports of agitations; (2) framework of the ambient environmental data processing; and (3) analysis techniques of the correlation between the environmental data and dementia-related agitation episodes. Figure 1 shows an overview of the proposed method.

### A. Data Collection

This work develops an in-home integrative sensing system that collects ambient environmental data in the homes of PWD [7]. The system consists of room-level sensing nodes which mounted with environmental sensors. Table 1 shows the list of environmental sensors and data captured by the sensing nodes and the sensing information. One sensing node is deployed in each room of the PWD homes, with exceptions of a privacy sensitive area such as bathrooms, to collect environmental data. To determine which sensing node is the closest to the PWD, we use a custom room-level localization system [8].

The system has been deployed to collect data in real PWD-CG dyad homes. We corroborate with clinical experts from the Carilion school of medicine, Virginia Tech. They provide us with feedback regarding PWD and CG interaction with the system and also recruiting PWD-CG dyad for real-world data collection. Each deployment takes two-months during which the system passively collects the environmental data without the need for CG attention to run the system. In this work, we used data from 3 data collection deployments in real PWD-CG dyad homes; each deployment covered 2-months' time period. During the deployment, CGs are encouraged to use a tablet survey application to record information about each PWD agitation episode, which includes: time of agitation, agitation severity, and agitated behavior of the PWD. The survey application is designed to required minimal burden from the CGs due to the already demanding tasks from taking care of the PWD. Furthermore, we design the system to make the data collection process unobtrusive and privacy protected such as avoiding the use of video cameras and eliminates any possible conversation captured by the audio sensor. Figure 2. shows an example of the collected ambient environmental data with an agitation label.

TABLE I. LIST OF ROOM-LEVEL ENVIRONMENTAL DATA COLLECTED BY THE SENSING NODES

| Environmental data | Sensors | Recorded unit | Data rate |
|---|---|---|---|
| Light | TAOS TSL2561 | Lux | 1Hz |
| Temperature | TI LM60 | ° Celsius | 1Hz |
| Humidity | Bosch BME280 | % relative humidity | 1Hz |
| Air pressure | Bosch BME280 | Barometric pressure (Pa) | 1Hz |
| Acoustic noise level | Omni-directional microphone | dB | 8Hz |

### B. Signal Processing Framework

*1) Data Filtering and Normalization:* Signal noises can be present in the data collected from environmental sensors mounted on the sensing nodes. The noise can be generated from the sensor interfacing channels or the sensitivity of the sensors. A 10-seconds median filter is used on all environmental data stream, except for the acoustic noise level, to reduce speckle noise. All environmental data values are normalized to the range of 0 to 100. The normalization is needed to balance the range of each environmental data. For example, the system can record air-pressure of 100,000 pascals while the recorded temperature value is 25-degree Celsius.

*2) Data Segmentation:* We extract 54-minutes of environmental data, from one hour to 12-minutes before the reported agitation episode. The extracted sequences of environmental data are segmented into 6-minutes windows without overlapping. This resulted in 9-windows per one agitation observation. We do not use the environmental data during the time of agitation in the segments, because we observed that there are certain changes in environmental data during most agitation episode. For example, we often see high-value of the acoustic noise level during the time of PWD agitation which can be explained as a verbally agitated behavior of the PWD or verbal activities between PWD and CG, as shown in figure 2. Since we are interested in the effect of the ambient environment on PWD which lead to agitation, we excluded the data at the time of agitation.

*3) Feature Extraction:* After the data is segmented into 9-sequences of 6-minutes windows (total of 54-minutes), features

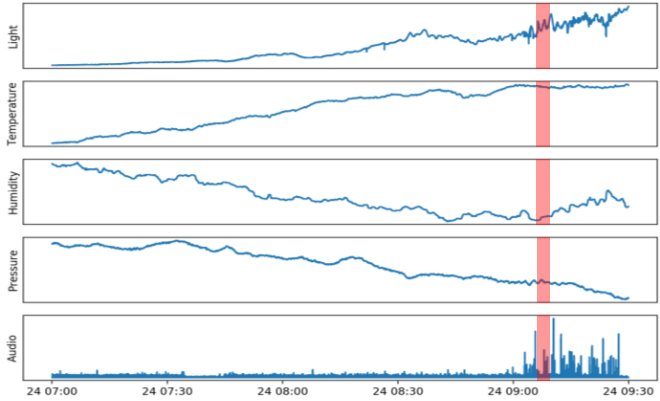

Fig. 2. A period of the ambient environmental data collected by a sensor node during a deployment in February 2019. The highlighted area shows an agitation label provided by the dementia caregiver.

are extracted from each of the windows. The features are used as an input to the prediction models. Features are extracted from the ambient environmental data to represents the current value of the data (e.g. if bright light level affects/causes dementia agitation). We also have a hypothesis that the sudden changes in the ambient environment can affect dementia agitation (e.g. if the rapid changes of light intensity affect/causes dementia agitation). Thus, we extract other feature sets that represent the changes in the environmental data. Additionally, we extracted the time-of-day feature as one of the model inputs. The time-of-day is hypothesized to be a useful agitation predictor in a case that dementia agitation often occurs at a similar time of the day [9]. The summary of all extracted features is shown in table 2.

TABLE II. LIST OF FEATURES EXTRACTED FROM EACH SIX-MINUTES WINDOW OF THE AMBIENT ENVIRONMENTAL DATA

| Type | Description | Features |
| --- | --- | --- |
| Data value | These features focus on the data value at the time within the six-minutes window | Mean |
| | | Median |
| | | Maximum |
| Data deviation | These features represent the changes in the environmental data within the window | Variance |
| | | Mean of differential |
| | | Maximum of differential |
| Other | - | Time-of-day |

### C. Prediction Models:

We compare two predictive machine learning models, gradient boosted trees (GBT) and long-short-term-memory (LSTM) neural network. The GBT model [10] is chosen because of the high observability that the model provides. Once the GBT model is trained, we can analyze the model's inputs using feature importance techniques [11] to determine which environment input contributes to predicting upcoming agitations. However, the GBT model is an instance-based model which considers input data as one instance of data points. This neglect the temporal relation in the ambient environmental data time-series.

To account for the temporal nature in data, the LSTM neural network is chosen as a comparison. The LSTM neural network is a recurrent neural network that can process the entire sequences of data, not just one data instance. LSTM model is structured with cells which remember outputs/predictions of the previous data instances in the data sequence and associate them in the later predictions in the time-sequences. We implemented the LSTM neural network to have 256 hidden nodes with 9 memory cells which represent the 9-time-windows extracted from an environmental data sequence.

Each model is trained exclusively on each deployment dataset and train on the combined dataset of the three deployments. This helps us to analyze whether the environmental triggers for agitation are specific to each PWD or not. We implemented a five-fold cross-validation technique to train/validate our models due to the small number of agitation observations (averaged 5 agitation reports per week). We use half of the observations as training/validating set and another half as the testing dataset. Non-agitation data sequences are randomly picked from the sensor station closest to the PWD when no agitation has been reported by the CG at the time. The nature of the real-world data, such as the reported agitation episodes, is highly imbalanced, e.g. fewer agitation episodes compare to non-agitation in real-world. Thus, the ratio between agitation observations and the randomly chosen non-agitation periods is one-to-three.

### III. RESULT

We evaluate the performance of the models using the agitation prediction accuracy, precision, and recall. Precision shows the ratio of correctly predicted dementia agitation to the total positive predicted agitation, and recall shows the ratio of correctly predicted agitation to all actual agitation reported. Figure 3 and 4 show the performance metrics of the GBT model and LSTM neural network model, in order. The GBT model yields 74% averaged accuracy, 64% averaged precision, and 65% averaged recall on the individual deployment datasets. LSTM shows better performance with 80% averaged accuracy, 70% averaged precision, and 79% averaged recall on the individual datasets. Another evaluation matric used to compare the model's performance is the weighted f1-score ($F_w$). The $F_w$ is chosen as a performance metric to account for the imbalance in the ratio between the number of agitation periods and the number of non-agitation periods. Table 3 compares the $F_w$ between models across training datasets.

The results show that the models trained on the individual datasets perform better than the ones trained on the combined dataset, in average, 10% better agitation prediction accuracy and 12% better $F_w$ for GBT model; and 17% better accuracy and 17% better $F_w$ for the LSTM neural network. LSTM also has a better performance in all matrics compared to the GBT model.

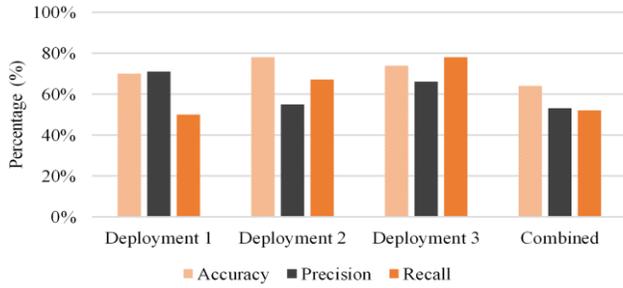

Fig. 3. Gradient boosted tree model performance for agitation prediction.

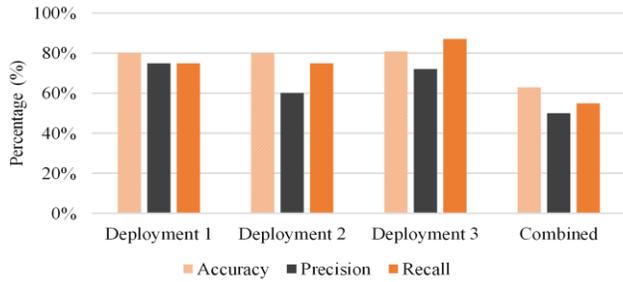

Fig. 4. Long-short-term-memory neural network model performance for agitation prediction.

This is likely due to the nature of the LSTM model which classify data based on a temporal relationship in the time-series data, such as the collected ambient environmental data.

TABLE III. WEIGHTED f1-SCORE OF EACH DEPLOYMENT FOR AGITATION PREDICTION

| Models | Datasets | | | |
|---|---|---|---|---|
| | *Deployment 1* | *Deployment 2* | *Deployment 3* | *Combined* |
| GBT | *0.70* | *0.79* | *0.74* | *0.62* |
| LSTM | *0.80* | *0.81* | *0.81* | *0.64* |

## IV. CONCLUSION AND DISCUSSION

This study focuses on the evaluation of using the ambient environment in PWD homes as predictors of upcoming agitation episodes. We collect in-home ambient environmental data in real-world deployments at the homes of PWD. The environmental data is processed and extracted as inputs to predictive machine learning models. The model is trained with periods of the reported agitation, provided by PWD caregivers. Results suggest that dementia agitation can be predicted from the in-home ambient environment with an average of 80% agitation prediction accuracy and 0.81 weighted f1-score for the LSTM neural network model. We also observe that the models trained on the individual deployment datasets perform better than the model with the combined dataset. This implies that the PWD's reaction to the ambient environment is dynamic and specific to each PWD.

The GBT model enables us to interpret the input feature importance which can help us understand the environmental triggers of dementia agitation. We analyzed the model's tree split gains and input weights. The split gain represents the inputs which have high contribution to the agitation prediction probability output of the model, and feature weight is the number of times that an input appears in the tree-based model, more appearance shows that the input is important to the model prediction output. However, none of the inputs stand out as a crucial predictor of the agitation. This could due to some limitations of the GBT model such as the neglection of time-series relationship in the data; the high number of inputs to the model (5 sensors, 7 features, 9-time windows, a total of 315 inputs); and the inaccurate timestamp of agitation occurrences. From this analysis, the only input feature that we observed to be highly correlated with the agitation occurrences is the time-of-day. This can be explained by the phenomenon called sundowning effect [12] which PWD is likely to become restless and agitated at a certain time of day, usually in the late afternoon or in the evening.

Overall, this work shows that the ambient environment can be monitored and used as predictors of dementia agitation. In the future, we plan to apply a more complex machine learning model interpretation techniques such as permutation importance to understand ambient environmental triggers of dementia agitations. Many clinical studies [13] reported that environmental treatments have shown to be effective in easing/intervening dementia agitations. We hope that by finding the environmental triggers of PWD agitation, we can construct effective agitation intervention/prevention strategies for individual PWD-CG dyad. Thus, reduce the CG burden associated with dementia agitations and extend aging in place.